\documentclass[12pt]{iopart}

\expandafter\let\csname equation*\endcsname\relax
\expandafter\let\csname endequation*\endcsname\relax

\usepackage{hyperref}
\usepackage{amssymb}
\usepackage[nobreak]{cite} 
\usepackage{physics} 
\usepackage{pm-isomath} 
\usepackage{todonotes}
\usepackage[utf8]{inputenc} 
\usepackage[T1]{fontenc}
\usepackage{bbm} 
\usepackage{siunitx}
\usepackage{upgreek} 
\usepackage[noabbrev]{cleveref} 
\crefname{equation}{}{}

\newcommand*{\aope}{\ensuremath{\hat a}} 
\newcommand*{\adag}{\ensuremath{\hat a^\dagger}} 
\newcommand*{\hc}{\ensuremath{\mathrm{H.c.~}}} 
\newcommand*{\avs}[1]{\ensuremath{\abs{#1}^2}} 

\newcommand*{\pii}{\ensuremath{\mathrm{\uppi}}} 
\newcommand*{\ii}{\ensuremath{\mathrm{i}}} 
\newcommand*{\eh}[1]{\ensuremath{\mathrm{e}^{#1}}} 

\newcommand*{\vs}[1]{\vectorsymbol{#1}} 
\newcommand*{\mas}[1]{\matrixsymbol{#1}}

\newcommand*{\refcite}[1]{~\cite{#1}}

\binoppenalty=\maxdimen
\relpenalty=\maxdimen

\begin{document}
\title[Time-dependent POVM reconstruction]{Time-dependent POVM reconstruction for single-photon avalanche photo diodes using adaptive regularization}
\author{E Fitzke, R Krebs, Th Haase, M Mengler\\ G Alber and Th Walther}
\address{Institute for Applied Physics, Technische Universität Darmstadt,\\Schlossgartenstra\ss e 7,  64289 Darmstadt, Germany}
\ead{thomas.walther@physik.tu-darmstadt.de}
\begin{abstract}
We report on the first realization of time-dependent quantum detector tomography for commercially available InGaAs avalanche photo detectors.
For the construction of appropriate time-dependent POVMs from experimentally measured data, we introduce a novel scheme
to calculate the weight of the regularization term based on the amount of measured data.
We compare our POVM-based results with the theoretical predictions of the previously developed model by Gouzien~et~al.~\cite{Gouzien2018}. In contrast to our measurement-based construction of a time-dependent POVM for photon detectors, this previous investigation extends a time-independent POVM to a time-dependent one by including effects of detector timing jitter and dead time on the basis of particular model assumptions concerning the inner physical mechanisms of a photon detector. Our experimental results demonstrate that this latter approach is not sufficient to completely describe the observable properties of our InGaAs avalanche photo detectors. 
Thus, constructing the time-dependent POVM of a detector by direct quantum tomographic measurements can reveal information about the detector's interior that may not easily be included in time-independent POVMs by a priori model assumptions. 
\end{abstract}


\maketitle
\section{Introduction}

Many applications in quantum information science such as the Boson sampling approach to quantum computing~\cite{Zhong_2021} or the characterization of quantum states~\cite{Auria_2009} can benefit from  detailed knowledge about the performance of single-photon detectors. Furthermore, the minimal requirements on detectors necessary for loophole-free Bell tests can be estimated when detector efficiencies as well as dark count rates are included in the analysis~\cite{Sauer2020}. Detailed knowledge about the detector's peculiarities can also be interesting in the context of quantum key distribution (QKD). QKD, proposed in 1984 by Bennett and Brassard, uses principles of quantum mechanics to distribute secure cryptographic keys~\cite{BB84, Gisin2002, Scarani2009}.
While in principle QKD provides information-theoretic security, actual implementations of QKD systems contain imperfections that can dilute this perfect degree of security.
Various attacks on single-photon detectors in QKD systems were demonstrated~\cite{Lydersen2010, Jogenfors_2015, Xu_2020}, stressing that detailed knowledge about the detectors is mandatory to maintain security. Alternatively, protocols immune to detector imperfections, known as measurement-device-independent QKD~\cite{Lo_2012}, with experimentally challenging requirements on the quality of the photon sources have to be employed~\cite{Wang2021}.

There are two fundamentally different approaches to detector characterization: The first approach is to thoroughly investigate all relevant effects on the measurement that arise from the detector's components and their interplay and to develop a detailed model of the detection process. However, this approach can easily become impractical for complex detector systems.
The second approach is quantum detector tomography, which aims to make as few assumptions as possible about the detector and instead reconstructs the measurement operator of a quantum detector from measurement results obtained from the detector itself~\cite{Luis1999,Feito2009,Lundeen2009}. Quantum detector tomography describes the detector by a positive operator-valued measure (POVM) completely characterizing the device.
The POVM can be reconstructed by analyzing the detection results obtained for a set of tomographically complete input states. Detector tomography can for example be used to characterize the qubit readout in quantum computers~\cite{Chen_2019}. 

A tomographically complete basis has to span the Hilbert space of the detector input states~\cite{Coldenstrodt2009}. So far, realizations of quantum detector tomography concentrate on single modes of the electromagnetic field, ignoring any time dependency~\cite{Lundeen2009}. Nevertheless, reduced tomographic analysis can yield valuable information about the figures of merit relevant in detector characterization, such as response time, dark count rate, efficiency, wavelength or photon-number resolution~\cite{vanEnk2017}.
Examples are tomographic measurements and POVM reconstruction for phase-insensitive detectors based on avalanche photo diodes~\cite{Feito2009}, time-multiplexed superconducting  detectors~\cite{Natarajan_13} or photon-number-discriminating nanostrip detectors~\cite{Endo_21} as well as analysis of phase-sensitive detectors~\cite{Zhang2012, Grandi_2017, Zhang_2012_Recursive}.

Although a time dependence is immanent to every measurement process, time-dependent tomography is rarely discussed. However, the detailed knowledge of the detector timing jitter is relevant for time-bin quantum measurements in QKD or quantum state tomography~\cite{Czerwinski_2021}, for example. A time-dependent theoretical model for POVMs of non-photon-number-resolving detectors including timing jitter and dead time was recently proposed by Gouzien~et~al.~\cite{Gouzien2018}.
Here, we extend the tomography of single-photon avalanche detectors to time-dependent POVMs and test the validity of the model proposed by Gouzien~et~al.\refcite{Gouzien2018} for our detectors. To the best of our knowledge this is the first experimental implementation  of time-dependent detector tomography. 

This paper is organized as follows:
First, we briefly review the theory of quantum detector tomography and introduce time-dependent POVMs in \cref{sec:theory}. By focusing on a single pulse shape for the input states we
reduce the dimensionality of the detector's input Hilbert space.
Since the reconstruction of the POVM elements from measured data is a mathematically ill-posed problem, regularization is necessary~\cite{Feito2009}. Often, the weight of this regularization is chosen by trial and error. Instead, we propose a novel adaptive regularization in \cref{sec:Adaptive_Regularizaton}, weighting the regularization based on the amount of measured data. We show benchmarking results of the proposed scheme in comparison with a fixed-weight regularization.
Our experimental setup is presented in \cref{sec:Setup}. With results integrated over the measurement time window we reconstruct the time-independent POVM of seven detectors, compare them with the expectation for ideal detectors with finite efficiency and deduce the detection efficiencies for different detector settings in \cref{ssec:Time-independent_POVMs}. 
Subsequently, we make use of the time resolution of the same data to reconstruct the time-dependent POVMs and apply them to one of the detectors in \cref{ssec:time-dependent_POVMs}. 
Finally, we compare our results in \cref{sec:timingjitter} with the model of Gouzien~et~al.~\cite{Gouzien2018} in order to evaluate its relevance for the theoretical description of our photon detector.

\section{Time-dependent quantum detector tomography in the photon number basis}\label{sec:theory}

In this section, we use a time-dependent detector POVM for describing a phase-independent click-or-no-click detector under the assumption that this detector is hit by non-entangled input states. This detector model is based on a model previously presented by Gouzien~et~al.~\cite{Gouzien2018} and takes advantage of a temporal multimode formalism as used by Rohde~et~al.~\cite{Rohde_2007}, for example. We generalize the previous work of Gouzien~et~al.~\cite{Gouzien2018} by not restricting ourselves to a specific model of the detector's inner working. We also briefly discuss the relationship between the POVM reconstruction of Feito~et~al.~\cite{Feito2009, Lundeen2009} and the maximum-likelihood estimation of the POVM elements used in the following.

The most general description of the measurement results of a quantum measurement process is given by a POVM~$\Pi$~\cite{Nielsen2010}, i.e. by a set of positive semi-definite measurement operators~$\Pi= \{\hat \Pi_i\}$ with~$\sum_i \hat \Pi_i = \mathbbm 1$, where~$i$ labels the different possible measurement results.  If a quantum state~$\hat \rho$ is prepared, the probability of obtaining measurement result~$i$ yields
\begin{equation}
\label{eq:genreal_POVM}
p_i(\hat\rho) = \tr(\hat\rho\hat\Pi_i)
\end{equation}
and~$p_i \geq 0$ is ensured by the positive semi-definiteness of the operators $\hat \Pi_i$~\cite{Luis1999}.
Quantum detector tomography is concerned with the reconstruction of these measurement operators from tomographic measurements~\cite{Feito2009}. 

Throughout this paper we consider detectors with two measurement results, i.e. 'click' and 'no click', in a time-dependent setting such that the probability density associated with a~'click' event at time~$t$ is given by
$p\ped{click}(t,\hat\rho) = \tr(\hat\rho\,\hat\pi\ped{click}(t))$.
The corresponding time-independent POVM for a time interval~$I$ is given by $\hat\Pi_{I,\,\text{click}} = \int_I \hat \pi\ped{click}(t) \dd t$ and~$\hat\Pi_{I,\,\text{no\,click}} = \mathbbm 1 - \hat\Pi_{I,\,\text{click}} $.
A complete time-dependent tomography is experimentally challenging, since it has to span the infinite dimensional space of all photon states at each instant of time~\cite{Rohde_2007}.
Thus, detector tomography is often reduced to a single mode~\cite{Feito2009, Lundeen2009, Natarajan_13, Endo_21} of the radiation field.

For a single mode, Fock states~$\ket{k}$ form a tomographically complete set of basis states. 
Here, we are interested in describing time-dependent phenomena. For this purpose we use the temporal multimode formalism from~\refcite{Rohde_2007} which has already been used to formulate a model for time-dependent POVMs by Gouzien~et~al.~\cite{Gouzien2018}. Thereby, we restrict the relevant Hilbert space to non-entangled time-localized states and assume that the detector dead time is much longer than the time interval considered so that at most one click can be registered in the time interval of interest.

Avalanche photo diodes do not have any external phase reference and are thus phase-insensitive detectors. POVMs of phase-insensitive detectors are diagonal in the Fock basis~\cite{Coldenstrodt2009} and can thus be described by POVM operators of the form
\begin{equation}
\label{eq:Fock_decomposition}
\hat\pi_i = \sum_{k=0}^\infty \Theta_{k,i} \op k.
\end{equation}
Including the time dependencies of the photons arriving at an avalanche photo diode we can represent the photon detector's POVM in the form
\begin{equation}\label{eq:pi_on_tau}
\hat \pi\ped{click}(t) = \mathcal{T} \sum_{k = 0}^\infty \int_{\mathbb R^k} p_{\text{click},\,k}(t, \vs\tau_k)\hat P_k(\vs \tau_k) \dd \vs \tau_k\,.
\end{equation}
Thereby,~$\vs \tau_k  = \{\tau_1,...,\tau_k\}$ denotes the times at which~$k$~time-localized photons arrive at the photon detector. The time-ordering operator is denoted by~$\mathcal T$ and ensures the ordering~$\tau_1<\tau_2< \dots < \tau_k$. The quantum state of the~$k$~time-localized photons is given by the projection operator
\begin{equation}\label{eq:Projection_Operator}
\hat P_k(\vs \tau_k) = \op{\vs \tau_k} \qq{with} \ket{\vs\tau_k} = \bigotimes_{j=1}^k  \hat{a}^{\dagger}(\tau_j)\ket{0}.
\end{equation}
\begin{sloppypar}
	Consistent with the rotating wave approximation, the creation operator $\hat{a}^{\dagger}(\tau_j) = (2\pii)^{-1/2}\int_{\mathbb R} \hat{a}^{\dagger}(\omega) \exp(\ii \omega \tau_j) \dd \omega$ describes the creation of a single time-localized photon at time~$\tau_j$ when the assumption is made that the bandwidth of the field excitation is much smaller than the optical center frequency~\cite{Loudon_2000, Brecht_2015}. The corresponding annihilation operators fulfill the commutation relation~$\comm{\hat a(\tau_i)}{\hat a^\dagger(\tau_j)} = \delta(\tau_i-\tau_j)$. The probability density~$p_{\text{click},\,k}(t, \vs \tau_k)$ describes the probability that a state with $k$~photons localized  at times~$\tau_1...\tau_k$ causes a click of the photon detector in the time interval~$[t,t+\dd t]$. 
	The POVM of \cref{eq:pi_on_tau} is a generalization of the time-dependent POVM model proposed by Gouzien~et~al.~\cite{Gouzien2018} as the probability density~$p_{\text{click},\,k}(t, \vs\tau_k)$ is not restricted to a specific model of the detectors's interior. In order to obtain a finite set of measurement operators, in the following we split the integral in~\cref{eq:pi_on_tau} into time bins of width~$\Delta t$ labeled from~$i =1$ to~$i\ped{max}$. Thus, the POVM to be reconstructed has $i\ped{max}+1$~different POVM operators, one for each time bin plus one for no click in any time bin.
\end{sloppypar}

Compared to Fock states the overcomplete basis of coherent states~$\ket{\alpha}$ is more convenient for describing experiments as coherent states are naturally produced by attenuated laser light. If a detector is exposed to~$N(\alpha_j)$ such attenuated light pulses of a coherent state $\ket{\alpha_j}$ the number of clicks concerning measurement result $i$, i.e.~$n_i(\alpha_j)$, can be recorded for all possible measurement results $i=1,\cdots,i\ped{max}$. This can be repeated for the different coherent states with $j=1,\ldots,j\ped{max}$. The resulting relative frequencies~$f_i(\alpha_j) = n_i(\alpha_j)/N(\alpha_j)$ can be compared to the probabilities~$p_i( \alpha_j)$ predicted by a given POVM.
In practice only a finite number~$j\ped{max}$ of different values of~$\alpha$ can be measured. As coherent states are linear superpositions of infinitely many photon number eigenstates, in practice also only a maximum number of photons~$k\ped{max}$ can be measured. Consequently, in terms of POVM parameters measured probabilities are described theoretically by the relation~$p_i(\alpha_j, \mas \Theta) = \sum_{k=0}^{k\ped{max}}\abs{\braket{\alpha_j}{k}}^2\Theta_{k,i}$. Thus, for a tomographic reconstruction of the POVM describing the photon detector the parameters~$\Theta_{k,i}$ have to be inferred from the measured frequencies $f_i(\alpha_j)$.

For the reconstruction the probabilities are approximated by the measured frequencies $f_i(\alpha_j)$, so that in matrix notation the relation between POVM elements and measured frequencies is given by
\begin{equation}\label{eq:Matrix_Equation}
\mas F_{j\ped{max} \times i\ped{max}} = \mas C_{j\ped{max} \times (k\ped{max}+1)} \mas \Theta_{(k\ped{max}+1) \times i\ped{max}}
\end{equation}
with~$C_{jk}= \abs{\braket{\alpha_j}{k}}^2 = \exp(-\mu_j)\mu_j^k/k!\,$ and with the mean photon number $\mu_j =\avs{\alpha_j}$.  

In general, the matrix~$\mas C$ is not invertible which complicates the determination of the POVM elements~$\Theta_{i,k}$ from the measured data in $\mas F$ with~$F_{ij} = f_i(\alpha_j)$. One possibility to solve this problem is to 
minimize $\norm{\mas F-\mas C\mas \Theta}\ped F$ with the Frobenius norm~$\norm{M}\ped F = (\sum_{i, j} \abs{m_{ij}}^2)^{1/2}$~\cite{Feito2009, Lundeen2009}.
In order to avoid unphysical solutions from this optimization problem it is convenient to add a regularization term to the objective function~\cite{pereyra2017maximum}. For example, in\refcite{Feito2009, Lundeen2009} a quadratic regularization term
\begin{equation}\label{eq:quadratic_regualrization_term}
r\sum_{k,i}(\Theta_{k+1,i}-\Theta_{k,i})^2
\end{equation} with a regularization coefficient~$r$ was used. The coefficient was chosen in a range so that a smooth distribution of the POVM elements is obtained and the reconstructed results are stable~\cite{Feito2009, Lundeen2009}.

In our subsequent treatment we split our measurement data into time bins and evaluate the bins individually. The number of recorded events per time bin varies across the considered time interval. The more data are available for a time bin, the smaller the statistical measurement uncertainty, which is reflected in the choice of~$r$. Instead of choosing a new value of~$r$ for each time bin by trial and error we propose an adaptive estimation value~$r$ which depends on the amount of measured data.

\section{Adaptive regularization}
\label{sec:Adaptive_Regularizaton}

In this section, we show the relationship between regularization terms that were used for POVM reconstructions in\refcite{Feito2009,Lundeen2009} and Bayesian prior distributions in the maximum-likelihood estimation of POVMs. We use this relationship to derive an estimation for the weighting coefficient~$r$ of the regularization term that depends on the amount of measured data and on the number of reconstructed elements. Subsequently, we benchmark the adaptive regularization scheme in comparison with regularization with a fixed coefficient.

In order to motivate an estimation value of~$r$ we consider the maximum-likelihood approach~\cite{Fiurasek_2001} for inversion of \cref{eq:Matrix_Equation}. The measured frequencies~$f_i(\alpha_j)$ may be viewed as the empirical mean values of a Bernoulli experiment with probabilities~$p_i(\alpha_j)$ and~$(1-p_i(\alpha_j))$ which has finite variance~$\sigma^2_{ij} = p_i(\alpha_j)(1-p_i(\alpha_j))$. The measured data are generated from statistically independent repetitions of Bernoulli experiments. For a sufficiently large number of repetitions, $\sigma^2_{ij} \approx f_i(\alpha_j)(1-f_i(\alpha_j))$ holds and the central limit theorem ensures that the distribution of~$f_i(\alpha_j)$ around $p_i(\alpha_j,\mas \Theta)$ is a normal distribution with variance $\sigma^2_{ij}/N(\alpha_j)$. This means the likelihood function is given by
\begin{equation}\label{eq:Likelihood_Gaussian}
L(\mas \Theta) = \frac{1}{(2\pii)^{i\ped{max}j\ped{max}/2}}  \exp(-\frac{1}{2}\sum_{i,j}\left(\frac{f_i(\alpha_j)-p_i(\alpha_j, \mas \Theta)}{N(\alpha_j)^{-1/2}\,\sigma_{ij}}\right)^2)\prod_{i,j}\sigma^{-1}_{ij} N(\alpha_j)^{1/2}\,.
\end{equation}

In order to obtain an estimation for the parameters $\mas \Theta$, the likelihood function can be maximized or, equivalently, the negative log-likelihood~$l(\mas \Theta) = -\ln(L(\mas \Theta))$ can be minimized. The sums over~$i$ are independent of each other, so that they can be minimized separately. When all constant factors are omitted and it is assumed that the~$\sigma_{ij}$ are independent of~$i$ for the same value of~$j$, minimizing the negative log-likelihood~$l(\mas \Theta)$ is equivalent to solving the least-squares minimization problem
$S(\mas \Theta) = \norm{\mas F-\mas C\mas \Theta}^2\ped F$.
The square root is a strictly monotone function, so that minimizing the norm $\norm{\mas F-\mas C\mas \Theta}\ped F$, as it was done in\refcite{Feito2009, Lundeen2009}, is also equivalent to the maximum-likelihood approach.

Adding regularization terms biases the optimization. The term of \cref{eq:quadratic_regualrization_term}, for example, biases POVMs towards close-by values for adjacent Fock basis matrix elements.
This bias can be interpreted as stemming from information predating the measurement in the form of a Bayesian prior distribution.  
Bayes' theorem allows to relate the likelihood~$P(\mas F|\mas \Theta)$ of detecting results~$\mas F$, given the parameters $\mas \Theta$, and the prior distribution~$P(\mas \Theta)$, to the posterior probability~$P(\mas \Theta|\mas F)$ of~$\mas \Theta$ being the parameter set if~$\mas F$ is measured: 
\begin{equation}\label{eq:bayes}
P(\mas \Theta|\mas F) = \frac{P(\mas F|\mas \Theta)P(\mas \Theta)}{P(\mas F)} \,.
\end{equation} 
Therefore, the (additive) regularization term can be understood as the negative log-likelihood of the (multiplicative) prior, so that the negative log-posterior function becomes
\begin{equation}\label{eq:log likelyhood_with_prior}
l\ped{posterior}(\mas \Theta) = l(\mas \Theta)-\ln(P(\mas \Theta)) + C
\end{equation}
with the additive constant~$C$ arising from the probability~$P(\mas F)$ of~\cref{eq:bayes}. 

Regularization terms as in \cref{eq:quadratic_regualrization_term} can thus be interpreted as a Gaussian prior~$P(\mas \Theta) = \exp(-\gamma\sum_{k,i}(\Theta_{k+1,i}-\Theta_{k,i})^2/2)$. 
The coefficient~$\gamma$ can be interpreted as the inverse covariance of neighboring matrix elements of $\mas \Theta$. Therefore,~$\gamma^{-1/2}$ is the expected characteristic distance of neighboring matrix elements. A prior can be constructed under the assumption that the matrix elements of $\mas \Theta$ are equidistantly spaced between~$0$ and~$k\ped{max}$. Consequently, it can be expected that the average distance between neighboring matrix elements is~$k\ped{max}^{-1}$ along the $k$-axis.  For the~$i\ped{max}$ different measurement results it can be assumed that they are equally distributed. This yields the relation~$\gamma=k\ped{max}^2 i\ped{max}^2$.

The coefficient~$r$ can also be related to the statistical measurement error which can be estimated by taking the maximum over the variances of
the normal distributions in \cref{eq:Likelihood_Gaussian} as 
\begin{equation}\label{eq:variance_from_beta_function}
\epsilon^2 = \max_{ij} \frac{\sigma^2_{ij}}{N(\alpha_j)} \approx \max_{ij} \frac{f_i(\alpha_j)(1-f_i(\alpha_j))}{N(\alpha_j)}\,.
\end{equation}
Here, the maximum is taken in order to obtain an upper bound on the uncertainty. 

The minimum of the negative log-posterior remains unchanged under multiplication with a positive constant. Therefore, we multiply the objective function by~$\epsilon^2$ and separate the regularization coefficient according to the relation~$r=\epsilon^2 \gamma$. Thus, for a fixed value of~$\gamma$, the weight of the regularization term becomes smaller for more accurate measurements. The more data are available the smaller the measurement uncertainty~$\epsilon^2$ and the more the prior bias is discounted.

In order to show the importance of regularization, especially if the amount of measured data becomes smaller, we now compare three different types of reconstruction schemes with least-squares minimization. These three schemes differ in the regularization used in the minimization. 
We compare results without regularization with static regularization, i.e. with a constant weight~$r$, and with the adaptive scheme motivated above where the regularization weight is adjusted by the variance of the measured data.
We used a value of~$r=0.1$ for the static case and a value of~$\gamma=k\ped{max}^2$ for the adaptive case, as we consider only one result, i.e.~$i\ped{max} = 1$. A cutoff of $k\ped{max}=29$ was chosen for the maximum Fock state used in the tomography and 30~coherent tomographic states were chosen with mean photon numbers $\mu=\{0,1,\dots,29\}$.
The data were recorded by randomly sampling measurement data for  known POVMs of detectors with only two results. Three different types of POVMs are studied in the benchmarking: two ideal detectors with sensitivities~$\eta=1$ and~$\eta=0.3$ with 'no-click' POVM elements~$\Theta_{0,k}=(1-\eta)^k$, and thirdly POVMs with randomly sampled diagonal elements. 

For the POVMs obtained from the reconstruction the maximum norm denoted by~$\ell_\infty = \max_k \abs{\theta_{k,\text{ true}}-\theta_{k,\text{ reconstr.}}}$ and the fidelity $\sum_k (\theta_{k,\text{ true}}\theta_{k,\text{ reconstr.}})^{1/2}$ with respect to the true POVM are compared. Within each simulated experiment all tomographic states were measured $M$~times. The same statistics is performed over various values of~$M$ in order to quantify the performance for an increasing amount of data. The simulation was repeated $N$~times for each value of~$M$ in order to estimate both the mean and the variance of the performance. 
The number of repetitions~$N$ was set to a value of~$100$.  For each POVM and each tomographic state $M$~measurements were sampled, a tomography was performed, and the results were compared with the true POVM.
Among the studied POVMs the diagonal elements of the random POVMs were newly sampled from a uniform distribution in each of the $N$~iterations. The other two ensembles used the same POVM across all iterations.

\begin{figure}[ht]
	\centering
	\includegraphics[height=0.5333\textwidth]{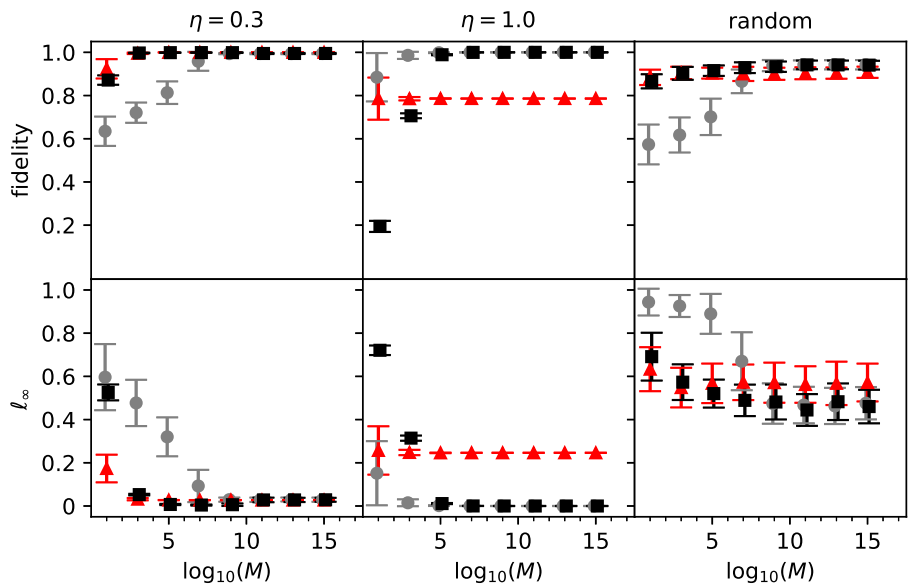}
	\caption{Benchmarking with detector type varying by column: The closeness of the POVM elements of the true detector and of the reconstructed elements are quantified by $\ell_\infty = \max_k \abs{\theta_{k,\text{ true}}-\theta_{k,\text{ reconstr.}}}$ and the fidelity $\sum_k (\theta_{k,\text{ true}}\theta_{k,\text{ reconstr.}})^{1/2}$. The first two columns show idealized detectors of finite sensitivity~$\eta$ without dark counts. Grey circles represent unregularized least squares, red triangles represent statically regularized least squares and black squares represent adaptive least squares. The $x$-axes show the number of trials~$M$ per tomographic state.} 
	\label{fig:BenchmarkingQuadratics}
\end{figure}
Results of this benchmarking are presented in \cref{fig:BenchmarkingQuadratics} clearly showing the improvement gained by regularization  for low values of~$M$ across all chosen underlying POVMs. For higher values of~$M$, all methods show improvement with increasing~$M$. For the random POVM, the data do not appear sufficient for high quality reconstruction, as both the error and its variance remain quite large, even for the highest values of~$M$, compared to the other two columns. Furthermore, it appears that the statically regularized method does not converge to optimum values, but to levels of a close-by but distinct value in all benchmarks. Only in the case of the detector with~$\eta=0.3$, all three methods seem to have similar asymptotic performance. The asymptotics for the statically regularized method emphasize that the regularization parameter should not be chosen independently of the number of data points.
In general, as expected from the vanishing regularization term, the asymptotic performance of the adaptively regularized method is very similar to that of \mbox{least-squares} tomography without regularization.
We thus use adaptive regularization for the reconstruction of POVMs in the following sections.

\section{Experimental setup}
\label{sec:Setup}

Tomography measurements were performed for seven free-running commercial InGaAs single-photon avalanche photo diodes (model ID220 with multimode fiber, ID Quantique). These detectors have three efficiency settings~(\SI{10}{\percent},~\SI{15}{\percent},~\SI{20}{\percent}) corresponding to different photo diode voltages. The dead time can be selected between \SIlist{1;20}{\micro\second}. In general, higher efficiencies and shorter dead times are preferable, but these settings come with a \mbox{trade-off}: the higher the efficiency is set, the higher the probability for detector afterpulsing  and dark counts for the same dead time setting. Afterpulses can be suppressed by choosing higher values for the dead time when a high efficiency is set. Thus, three combinations of detector settings were chosen for the experiments: \SI{5}{\micro\second} dead time for \SI{10}{\percent} set efficiency, \SI{10}{\micro\second} for \SI{15}{\percent} and \SI{15}{\micro\second} for \SI{20}{\percent}.

\begin{figure}
	\centering
	\includegraphics[width=0.66\textwidth]{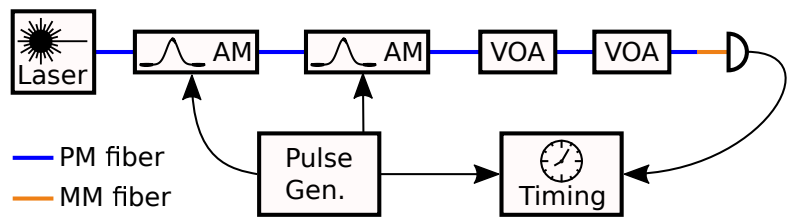}
	\caption{Fiber-based setup for time-dependent and time-independent detector tomography. \emph{AM}:~LiNbO\textsubscript{3} amplitude modulator, \emph{VOA}:~Variable optical attenuator, \emph{Pulse~Gen.}:~Electronic pulse generator,
		\emph{Timing}:~Time tagging electronics, \emph{PM~fiber}:~Polarization maintaining fiber, \emph{MM~fiber}: Multimode fiber}
	\label{fig:setup}
\end{figure}

In order to perform both time-independent and time-dependent tomography, we set up the experiment shown in \cref{fig:setup}. Laser pulses with a defined mean photon number~$\mu$ were generated and detector clicks were registered correlated to the pulse emission.

The setup consists of a fiber-coupled cw DFB diode laser with a central wavelength of \SI{1550.52}{\nano\meter} and \SI{74}{\milli\watt} output power, two cascaded amplitude modulators, a manual and an electronic variable attenuator as well as electronics for timing acquisition and pulse generation. 

The first modulator was used to shape pulses with a FWHM duration of \SI{0.24}{\nano\second} at a repetition rate of \SI{10}{\kilo\hertz}. The rate was chosen to be low enough so that a repetition cycle was much longer than the detector dead time. Hence, unwanted correlations between subsequent clicks introduced by afterpulses or variations in the dead time were prevented. 
The pulse duration for the second modulator was set to \SI{10}{\nano\second} and the delay was set according to the optical delay between both modulators, ensuring that the second modulator was completely opened when a pulse passed by. Therefore, the pulse shape was solely determined by the first modulator. Within the opening time window of the second modulator, the extinction ratio is determined by the first modulator only. Outside of the time window, the extinction ratios of both modulators multiply, resulting in a sufficiently high suppression of uncorrelated detection events from photons leaking through the modulators during the time between pulses.

Both modulators were driven by a dual-channel pulse generator (HP 8131 A). The detector's electrical output as well as the trigger output of the pulse generator were connected to the timing acquisition  electronics (ID Quantique ID900) with \SI{13}{\pico\second} resolution. Timestamps for detector clicks and the trigger pulses were recorded in different channels.

The first modulator was stabilized by a bias controller in order to prevent changes in the pulse shape and energy. In order to avoid bias drifts of the second modulator, a recalibration of the bias voltage was run before measuring each value of the mean photon number~$\mu$ by sweeping the voltage and setting the bias voltage that minimized the rate of uncorrelated counts.

For the tomography measurements, the average power was measured behind the second modulator and the pulse energy was calculated from the repetition rate. The mean photon number of the pulses~$\mu$ was scanned by adjusting the attenuation value of the variable attenuator. 

\section{Reconstruction of time-independent POVMs}
\label{ssec:Time-independent_POVMs}

In order to calculate time-independent POVMs for the detectors, histograms for the time difference between reference clicks from the pulse generator and detector clicks were calculated for a measurement series over the mean photon numbers~$\mu$ between 0 and~$\mu\ped{max} = \num{50}$ in steps of~2 with \SI{10}{\minute} measurement time per value. In order to avoid effects from the detector dead time and afterpulsing, clicks were excluded which were preceded by another click within a time frame of the set dead time plus two microseconds. The detection probabilities were calculated by dividing the sum of detections within a time window of \SI{8}{\nano\second} around the maximum of the histogram by the number of pulses. In this way, we obtained the time-independent POVMs from the detection probabilities by minimizing
\begin{equation}\label{eq:POVM_reconstruction_formula}
\norm{\vs f-\mas C \vs \theta }^2_2 + \epsilon^2 \gamma \sum_{k=0}^{k\ped{max}} \left(\theta_{k+1}-\theta_{k}\right)^2
\end{equation}
over~$\vs \theta$. Here, the matrices~$\mas F$ and~$\mas \Theta$ became vectors~$\vs f$ and $\vs \theta$, as only one single detection result, the no-click event in the whole interval, was considered.

For the photon numbers, a reasonable cutoff~$k\ped{max}$ needs to be found. The coefficients~$C_{jk}$ decrease for higher~$k$ according to the Poissonian distribution~$\exp(-\mu) \mu^k/k!$ with the standard deviation of the photon number given by~$\mu^{1/2}$. We decided to reconstruct the elements up to~$\mu\ped{max}$ plus two standard deviations, i.e. chose~$k\ped{max}  \approx \mu\ped{max} + 2 \sqrt{\mu\ped{max}} = 64$ resulting in~$C_{j\ped{max}k\ped{max}} < \SI{1}{\percent}$. Thus, 
the weighting coefficients of the regularization term are~$\gamma = 64^2$ and~$\epsilon^2 \approx \num{4.4e-8}$ which was calculated for detector~2 with \SI{15}{\percent} efficiency and \SI{10}{\micro\second} dead time according to
\cref{eq:variance_from_beta_function}. Convergence of the minimization was facilitated by explicitly implementing the gradient of \cref{eq:POVM_reconstruction_formula}.

In order to compare the results, we consider the POVM for an ideal detector with efficiency~$\eta$. When such a detector is exposed to a photon, the probability that it is not triggered is given by~$(1-\eta)$. Thus, the no-click probability for $k$~photons is given by~$P\ped{no\:click}(k) = 1- P\ped{click}(k) = (1-\eta)^k$ and  \cref{eq:Fock_decomposition} becomes~$\hat \pi\ped{no\:click} = \sum_{k=0}^{\infty}(1-\eta)^k \op{k}$.
Thus, the no-click probabilities for a number state~$\ket k$ and for a coherent state~$\ket \alpha$ of an ideal detector are given by
\begin{equation}\label{eq:time-indpendent_noclick_probability}
\ev{\hat \pi\ped{no\:click}}{k} = (1-\eta)^k \qand \ev{\hat \pi\ped{no\:click}}{\alpha} = \exp(-\eta \mu)\,.
\end{equation}
In order to obtain a value for the detector efficiency, the POVM element for~$k=1$ can be considered or, alternatively, the efficiency can directly be derived from the measured data by fitting an exponential function to the number of clicks over the mean photon number~$\mu$ according to \cref{eq:time-indpendent_noclick_probability}.

\begin{figure}[ht]
	\centering
	\includegraphics[height=0.4\textwidth]{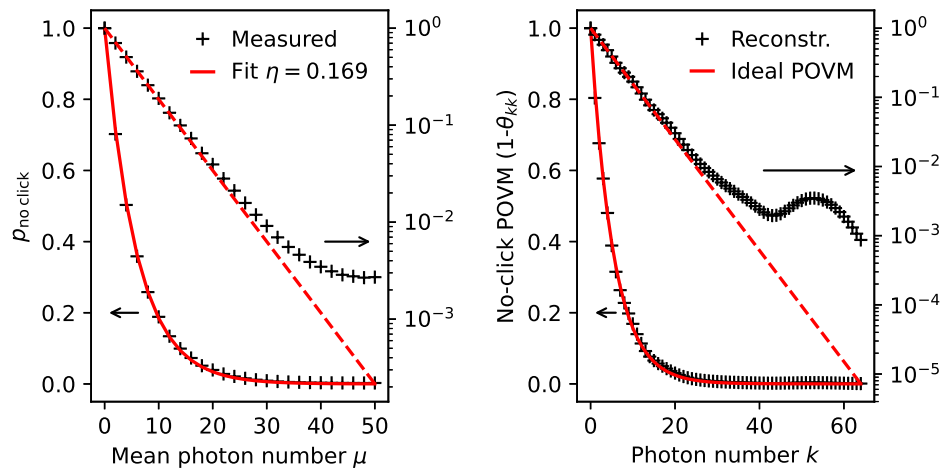}
	\caption{Time-independent POVM elements for detector~2 (cf.~\cref{fig:efficiencies}) with \SI{15}{\percent} efficiency and \SI{10}{\micro\second} dead time: From a fit of the no-click probabilities according to \cref{eq:time-indpendent_noclick_probability}, the detector efficiency was estimated to~$\eta=\SI{16.9}{\percent}$. This value was used to plot the ideal no-click POVM according to \cref{eq:time-indpendent_noclick_probability} on the right-hand side. Each diagram shows the data once in a linear scale (left vertical axis) and in a logarithmic scale (right vertical axis, dashed line).}
	\label{fig:time-independent_POVMs}
\end{figure}

\begin{figure}[ht]
	\centering
	\includegraphics[height=0.4\textwidth]{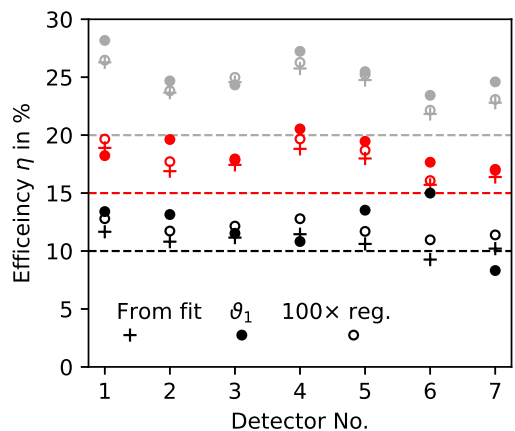}
	\caption{Detection efficiencies of the seven detectors tested, calculated from an exponential fit to~$p\ped{no\,click}(\mu)$ and from the first POVM element $\theta_1$, reconstructed with the proposed weighting factor~$r=\epsilon^2\gamma$ of the regularization term: In addition, results with a 100~times stronger regularization are shown. The colors indicate the detector efficiency settings \SI{10}{\percent} (black), \SI{15}{\percent} (red) and \SI{20}{\percent} (gray).
	}
	\label{fig:efficiencies}
\end{figure}

The time-independent POVMs were reconstructed for all seven detectors. Exemplary results for detection probabilities and time-independent POVMs of detector~2 (cf.~\cref{fig:efficiencies}) with an efficiency setting of \SI{15}{\percent} and \SI{10}{\us} dead time are shown in \cref{fig:time-independent_POVMs} along with a fit for an ideal detector. 
The measured no-click probability well matches the exponential distribution of an ideal detector with~$\eta = \SI{16.9}{\percent}$ according to \cref{eq:time-indpendent_noclick_probability}. For values above~$\mu = 30$, however, the logarithmic scale shows that the  no-click probability is higher than for an ideal detector and approaches a value of \num{2.7e-3}, irrespective of a further increase of~$\mu$.
In order to investigate this effect, the selection criterion for clicks was extended from \SIrange{12}{99}{\micro\second} as the required distance to the preceding click. However, no change was observed. Thus, we conclude that this effect is independent of the time difference to the previous click and is not related to the dead time.
The POVM elements shown in \cref{fig:time-independent_POVMs} reflect this behavior: Up to photon numbers around 30, they match the model of an ideal detector well, but for higher values of~$k$ they are larger than predicted by the model.

For all seven detectors the efficiencies obtained from the POVM reconstruction are shown in \cref{fig:efficiencies}.
Systematic relative measurement uncertainties for~$\mu$ are introduced by the accuracy of the photo detector used for attenuator calibration (\SI{5}{\percent}) and by variations of losses in the fiber-fiber connections (\SI{10}{\percent}). However, these values are constant for the measurements shown. In principle, these values can be further improved by using a tightly calibrated photo detector for the attenuation calibration and by using permanent, spliced fiber connections. Repeated measurements of the same detector yielded a variation of \SI{8}{\percent}. 

We determined the values for the detection efficiencies in three different ways which are compared in the figure. First, the efficiency was extracted from a fit of the exponential distribution in \cref{eq:time-indpendent_noclick_probability} to the measured values of~$p\ped{no\,click}(\mu)$. Second, the POVM elements were reconstructed with the proposed adaptive weight $r=\epsilon^2\gamma$ of the regularization term and the value~$(1-\theta_1)$ was interpreted as efficiency. Third, the second procedure was repeated with a 100 times stronger regularization.

The determined efficiency values in \cref{fig:efficiencies} match the expected values stated as detector settings generally being slightly higher than these values. For most of the detectors the efficiency obtained from the strongly regularized reconstruction is in better agreement with the value determined from the fit than the value obtained with normal regularization. The fact that reasonable POVMs can be obtained although the regularization coefficient can be varied by more than two orders of magnitude has already been observed in\refcite{Feito2009}. In this previous investigation it has been concluded that the regularization is mainly necessary to ensure a well-conditioned optimization and that choosing the regularization coefficient in this range does not excessively distort the results. Thus, we conclude that the proposed adaptive regularization coefficient can be understood as a rule-of-thumb value to obtain reasonable results from the POVM reconstruction. Notably, it is not a strict value so that larger or smaller values may also be chosen, depending on the specific situation.

\section{Time-dependent POVMs}
\label{ssec:time-dependent_POVMs}
\begin{figure}[ht]
	\centering
	\includegraphics[height=0.4\textwidth]{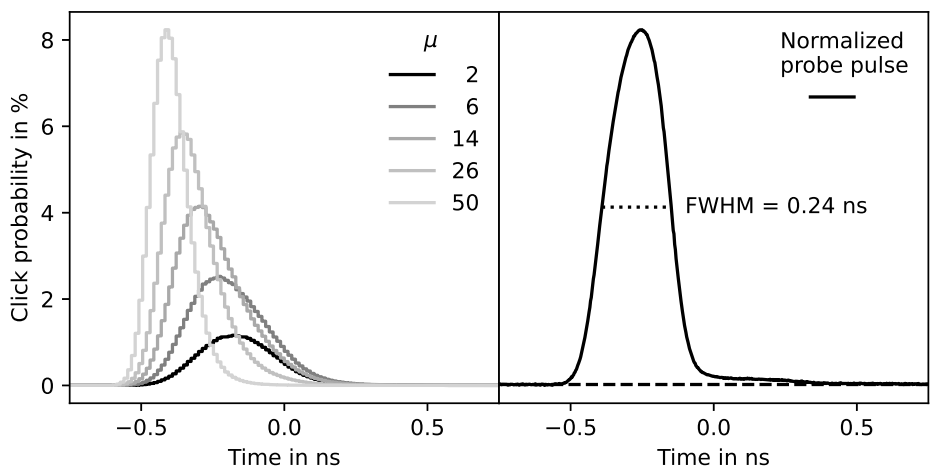}
	\caption{Click probabilities~$p\ped{wp}(t)\Delta t$ of detector~2 (cf.~\cref{fig:efficiencies}) with a setting of \SI{15}{\percent} efficiency and \SI{10}{\us} dead time: The time resolution is~$\Delta t = \SI{13}{\pico\second}$. On the right hand side the normalized probe pulse shape~$\abs{\alpha(t)}^2$ is shown for comparison. The absolute delay between the pulse and the click distribution depends on optical and electrical delays. Both time axes were shifted so that the pulse shape and the click distribution are located close to time~$t=0$.} 
	\label{fig:measured_rates}
\end{figure}

The temporal resolution of the setup also allowed for a time-resolved measurement of detection probabilities. In general, the temporal distribution of the clicks depends on the temporal shape of the probe pulse and on the detector response. Exemplary results for detector~2 are shown in \cref{fig:measured_rates} along with the probe pulse shape.
With increasing values of~$\mu$, the maximum of the click distribution shifts to earlier times with respect to the input pulse and also becomes narrower. This effect can be understood intuitively: After the click the detector switches into the dead time, so that other photons in the pulse can not cause subsequent clicks for the same pulse. The higher~$\mu$, the higher the probability that a photon located early within the pulse causes a click. The deformation of the click distribution can be expected to become significant above~$\eta\mu\approx 1$, as a detector without dead time would likely yield multiple clicks per pulse, but a detector with dead time only registers the first click per pulse.

\begin{figure}[ht]
	\centering
	\includegraphics[height=0.4\textwidth]{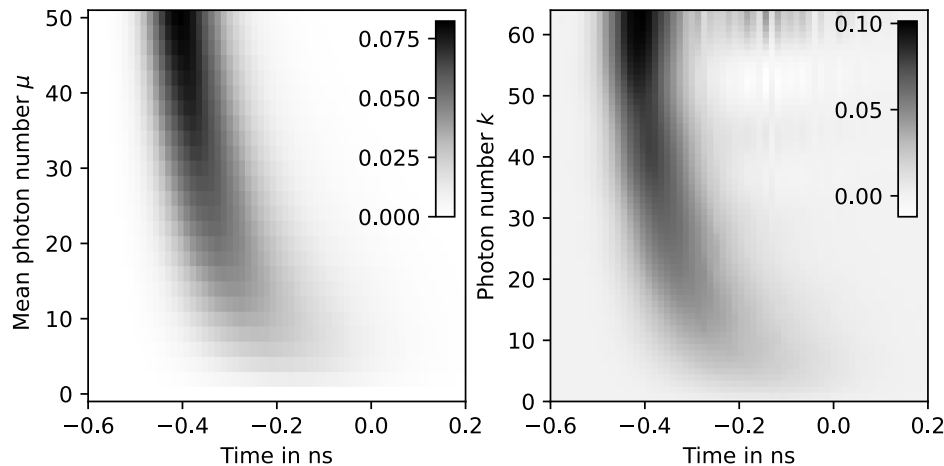}
	\caption{Left:~Measured time-dependent click distribution depicted as a function of the mean photon number~$\mu$ for detector~2 (cf.~\cref{fig:efficiencies}) with \SI{15}{\percent} set efficiency and \SI{10}{\us} dead time. Right:~Time-dependent POVM elements are reconstructed with adaptive regularization from the click distribution.}
	\label{fig:Time-dependent_POVMs_2D}
\end{figure}

From the resulting click probability distribution, a time-dependent POVM as in \cref{eq:Fock_decomposition} was  reconstructed, where clicks in a specific time bin correspond to a specific detection result~$i$. Here, we minimized \cref{eq:POVM_reconstruction_formula} for each time bin individually, with the variance $\epsilon^2$ calculated from the data in this time bin only. The measured click probabilities and the reconstructed POVM for detector~2 are shown in  \cref{fig:Time-dependent_POVMs_2D}.

Again, the POVM was reconstructed up to~$k = 64$. In the region~$k \geq 58$, numerical artifacts begin to arise, so that a reconstruction which also takes into account higher values of~$k$ is not useful.
The shape of the POVM roughly matches that of the click probability distribution. This raises the question whether the description of the detector by a time-dependent POVM is necessary or if the temporal shape of the distribution can be explained by modeling the detector behavior. In the next section, we compare the reconstructed POVM to the POVM predicted by a detector model describing the click probability deformation explained above. 

\section{Test of a detector model including dead time and timing jitter}
\label{sec:timingjitter}

In the last section we argued that the changes of the shape of the click distribution with increasing mean photon number $\mu$ qualitatively match the expectation for a detector which only responds to the first photon in each pulse. Gouzien~et~al.~\cite{Gouzien2018} formalized this argument and developed a model describing time-dependent POVMs of click-or-no-click detectors. In this section, we investigate whether this model is sufficient in order to describe the observed temporal detector click probability for our detectors.

The model of Gouzien~et~al.~\cite{Gouzien2018} assumes a detector having an intrinsic jitter distribution~$J(T)$. The value~$J(t-\tau)\dd t$ is the probability that a photon hitting the detector at time~$\tau$ causes a click in the time interval~$[t, t+\dd t]$. Causality requires that~$J(t<0) = 0$. Furthermore, it is assumed that the dead time is much longer than the probe pulse duration and that the detector switches immediately into dead time after the first click. This means that after a click the detector is inactive for the remaining pulse duration and all other photons in the pulse cannot cause subsequent clicks.
The probabilities
\begin{equation}\label{eq:p_tau}
p_1(t, \tau) = \eta J(t-\tau) \qand p_{1, \text{not}}(t, \tau) =  1- \eta \int_{\tau}^{t}J(t'-\tau) \dd t'
\end{equation}
describe the probability that a single photon hitting the detector at~$\tau$ causes a click at~$t$ and that a photon has not caused a click up to time~$t$, respectively. 

The probability~$p_{\text{click}}(t, \vs \tau_k) $ of obtaining a click at time~$t$ from a pulse with~$k$~photons arriving at times~$\vs \tau_k  = (\tau_1,...,\tau_k)$ can now be written as the sum of the probabilities that one particular photon causes the click and the probability that all other $k-1$~photons did not yet cause a click:
\begin{equation}\label{eq:p_on_tau}
p_{\text{click}}(t,\vs \tau_k) = \sum_{j=1}^{k}p_1(t,\tau_j)\prod^k_{\substack{l=1 \\l\neq j}}p_{1, \text{not}}(t,\tau_l)
\end{equation}
Multiplying the probabilities is justified by the assumption that apart from the dead time effect the photons cause clicks according to the jitter distribution independently of each other. The POVM model proposed in\refcite{Gouzien2018} is a combination of the general time-dependent POVM from \cref{eq:pi_on_tau} and the specific form of $p_{\text{click}}(t,\vs \tau_k)$ from \cref{eq:p_on_tau}. In\refcite{Gouzien2018} the click distributions for single-photon and two-mode biphoton states are calculated for this specific POVM.

In order to compare the predictions of this model with our experimental results we calculate the click distribution for a coherent wavepacket. Therefore, we insert \cref{eq:p_on_tau} into the general formula for the time-dependent POVMs \cref{eq:pi_on_tau} and apply it to a continuous multi-mode coherent wave packet 
\begin{equation}\label{eq:wave_packet_input state}
\ket{\Psi\ped{wp}} = \exp(\int_{\mathbb{R}} \left(\alpha(t)\adag(t)- \hc \right) \dd t) \ket 0.
\end{equation}
It is assumed that the wave packet has a sufficiently narrow bandwidth and~$[\aope(t), \adag(t')] = \delta(t-t')$ holds~\cite{Loudon_2000}.
Thus,~$\avs{\alpha(t)}$ describes the time-dependent photon flux of the probe pulse with mean photon number~$\mu = \int_{\mathbb R}\avs{\alpha(t)}\dd t$. 
In the \hyperref[appendix]{appendix} we show that the probability of obtaining a click at time~$t$ for such a wave packet~$\ket{\Psi\ped{wp}}$ is given by
\begin{equation}\label{eq:Gouzien_applied_to_wavepacket}
p\ped{wp}(t) = -\pdv{}{t} \exp(-\eta  \int_{-\infty}^{t}(J*\abs{\alpha}^2)(t')\dd t')\,, 
\end{equation}
with $*$ indicating convolution.

This equation has a structure known from Poisson processes. For an inhomogeneous Poisson process with time-dependent rate $\lambda(t)$, the probability for the first detection in the interval~$[t, t+\dd t]$ is~$p_\lambda(t)\dd t = \lambda(t)\text{exp}\big(-\int_{-\infty}^t\lambda(t') \dd t'\big) \dd t$~\cite{Streit2010}.
Thus,~$p\ped{wp}(t)$~resembles the probability density for the time up to the first click of an inhomogeneous Poisson process with the  click rate 
\begin{equation}\label{eq:lambda_rate_definition}
\lambda(t)= \eta(J*\abs{\alpha}^2)(t)\,.
\end{equation}
The structure of~$p\ped{wp}(t)$ can be understood by recalling that the detection of a coherent state yields Poissonian statistics. Here, the temporal shape of the wave packet is modified by convolution with the intrinsic detector jitter distribution. The Poissonian statistics reflects the fact that all photons are treated independently. As only the first click is registered due to the detector switching into dead time, the resulting distribution is the probability density of the first-click-time of this process.

\begin{figure}[ht]
	\centering
	\includegraphics[height=0.4\textwidth]{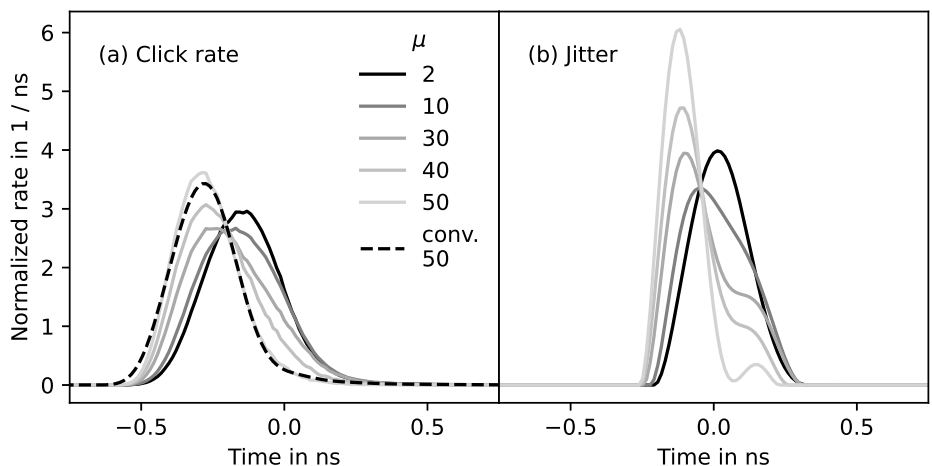}
	\caption{Investigation of the model from\refcite{Gouzien2018} for detector~2 (cf.~\cref{fig:efficiencies}) with \SI{15}{\percent} efficiency and \SI{10}{\us} dead time: (a)~Normalized click rate~$\lambda(t)/(\eta\mu)$ according to \cref{eq:clickrate}: The dashed line shows a \mbox{re-convolution} of the obtained jitter distribution for~$\mu = 50$ with the pulse shape. (b)~Reconstructed intrinsic jitter distributions~$J(T)$ obtained from deconvolution of \cref{eq:lambda_rate_definition}.}
	\label{fig:Jitter_Distributions}
\end{figure}

In order to check whether this model is valid for our detectors, we investigated whether it is possible to reconstruct the jitter distribution~$J(T)$ from the measured click distribution~$p\ped{wp} (t)$ according to \cref{eq:Gouzien_applied_to_wavepacket}.
In order to reconstruct~$J(T)$, we define the cumulative rate~$\Lambda(t) = \int_{-\infty}^t \lambda(t')\dd t' $ and write~$\int_{-\infty}^t p\ped{wp}(t')\dd t' = 1- \exp(-\Lambda(t))$, where we used~$\Lambda(t\rightarrow -\infty) = 0$. Solving for~$\lambda(t)$ yields 
\begin{equation}\label{eq:clickrate}
\lambda(t) = \pdv{\Lambda(t)}{t} = -\pdv{t}\ln(1-\int_{-\infty}^t p\ped{wp}(t')\dd t') =  p\ped{wp}(t)\left(1-\int_{-\infty}^t p\ped{wp}(t)\right)^{-1}\,.
\end{equation}
The right hand side can be directly calculated from the measured data, without requiring a calculation of $\eta$ or $\mu\ped{wp}$. The click rates~$\lambda(t)$ are shown in \cref{fig:Jitter_Distributions} for different values of~$\mu$.

We also computed the jitter distribution~$J(T)$ by deconvolution from~$\lambda (t)$ according to \cref{eq:lambda_rate_definition} for different mean photon numbers~$\mu$. 
For the numerical implementation of the deconvolution with discrete data, it is convenient to introduce the normalized discrete pulse shape~$\vs I$ with~$I_i = \abs{\alpha(t_i)}^2/\mu\ped{wp}$ and the normalized rate~$\vs  \lambda$ with~$\lambda_i = \lambda(t_i)/(\mu\ped{wp}\eta)$, with~$\sum_i \lambda_i = \sum_i I_i = 1$. The discrete convolution with the jitter~$J_i$ can then be written as~$\vs \lambda = (\vs I* \vs J) = \mas T \vs J$, where~$\mas T$ is a Toeplitz matrix constructed from~$\vs I$. The deconvolution can be performed by minimizing~$\norm{\vs \lambda - \mas T \vs J}^2_2$ over~$\vs J$. The formulation with the Toeplitz matrix enabled direct implementation of the gradient by using~$\nabla_J \norm{\vs \lambda - \mas T \vs J}^2_2 = 2\mas T^\mathsf{T}(\mas T \vs J - \vs \lambda)$, which facilitated the convergence. We optimized with the constraint $J(T)\geq0$. In order to obtain smooth results, we penalized large variations in the first derivative of the jitter distribution by introducing the regularization term~$\sum_{i}(J_{i+1}-J_i)^2$, multiplied by a weighting coefficient. The objective function is thus very similar to the objective function for the POVM reconstruction in \cref{eq:POVM_reconstruction_formula}.

According to the model, the detector should have one distinct jitter distribution that explains the resulting click distributions for all values of~$\mu$ according to \cref{eq:Gouzien_applied_to_wavepacket}. Consequently, the normalized rate~$\lambda(t)$ and the deconvolved jitter distribution should be independent of~$\mu$. However, with increasing~$\mu$, the rate distribution becomes narrower and shifts towards earlier times, meaning that the model underestimates the previously discussed deformation of the click probabilities in \cref{fig:measured_rates}. The effect is even more pronounced in the jitter distributions and appeared for all seven detectors and for all three  detector settings.
For high values of~$\mu$, a foothill appears in the jitter distribution in \cref{fig:Jitter_Distributions} and becomes a side maximum for~$\mu = 50$. A \mbox{re-convolution} of this jitter distribution with the pulse shape shows that for~$\mu = 50$ the top of the peak is not accurately described any more by the convolution, meaning that there is no jitter distribution which, when convolved with the pulse shape, results in this particular shape of~$\lambda(t)$. From these two observations we conclude that the model is not sufficient to completely describe the time-dependent click distribution of our detectors.
A possible explanation for the deviation can be found in the detection mechanism by electron avalanches. When the amplitude of an electron avalanche reaches a threshold level, the detector emits an electric pulse to indicate a detection event. The main contribution of the timing jitter in InGaAs single-photon avalanche detectors comes from the distribution of transit times of charge carriers in the absorption region and by the distribution of the avalanche build-up time in the multiplication region~\cite{Amri_2016}. When multiple photons hit the detector, the avalanches can add up. The threshold for a detection event is thus reached faster than expected for photons triggering independent events. It can be expected that the effect will cause a shift of the click distribution to earlier times that is stronger than the prediction of the model. The underlying general reason may thus be the violation of the model assumption that, except for the dead time effect, all photons can be treated independently. An option to include such effects would be to develop complex models of the detector behavior including more details of the detection mechanism such as in\refcite{Tan_2007, Petticrew_2018}. However, this approach is contrary to the tomographic approach, which is to introduce as few as possible general assumptions about the detector.

\section{Conclusion}

We performed tomographic measurements on avalanche single-photon detectors that enabled the reconstruction of both time-independent and time-dependent POVMs. The time-independent tomography results were in agreement with a simple model of an ideal detector. By this method, we deduced and compared the detection efficiencies of seven detectors. For the time-dependent POVM reconstruction we derived an estimation for the weighting coefficient which adapts the regularization term based on the amount of available data in different time bins. Benchmarking the new method showed a superior performance in comparison with reconstructions based on a fixed coefficient. We then reconstructed time-dependent POVMs by using the adaptive regularization. Finally, we investigated whether the model for time-dependent POVMs proposed by Gouzien~et~al.\refcite{Gouzien2018} can explain the measured POVMs and showed that the model 
is not able to explain the performance of our photon InGaAs single-photon avalanche detectors in a satisfactory way. This example demonstrates the strength of detector tomography in comparison with less flexible modeling approaches. Thus, measuring the time-dependent POVM of a detector with quantum tomographic methods can reveal information about the detector's interior that may not easily be included by a priori model assumptions. 

\section*{Acknowledgement and Data Availability}
This research has been funded by the Deutsche Forschungsgemeinschaft (DFG, German Research Foundation) – SFB 1119 – 236615297

Data available on request from the authors.

\section*{References}
\providecommand{\newblock}{}

\clearpage
\appendix
\setcounter{section}{1}
\section*{Appendix}
\label{appendix}

Here, we calculate the click probability~$p\ped{wp}(t)$ for Gouziens's model~\cite{Gouzien2018} applied to a coherent wavepacket by inserting \cref{eq:p_on_tau} into \cref{eq:pi_on_tau}
\begin{equation}
p\ped{wp}(t) =\ev{\hat \pi\ped{click}(t)}{\Psi\ped{wp}}= \mathcal T \sum_{k = 0}^\infty \int_{\mathbb R^k} p_{\text{click}}(t, \vs \tau_k)  \ev{\hat P_k(\vs\tau_k)}{\Psi\ped{wp}}  \dd \vs \tau_k.
\end{equation}
The integral is subject to time ordering~$\tau_1<\tau_2< \dots < \tau_k$. However, both~$p\ped{click}(t,\vs\tau_k)$ from \cref{eq:p_on_tau} and~$\ev{\hat P_k(\vs \tau_k)}{\Psi\ped{wp}}$ from \cref{eq:Projection_Operator} are symmetric under permutation of~$\tau_1... \tau_k$. The time ordering can thus be expressed by extending the integration range to the complete real line for each~$\tau$ and simultaneously dividing by the number of~$k!$ permutations
\begin{equation}
p\ped{wp}(t) = \sum_{k = 0}^\infty \frac{1}{k!}\int_{\mathbb R^k} p_{\text{click}}(t, \vs \tau_k)  \ev{\hat P_k(\vs\tau_k)}{\Psi\ped{wp}}  \dd \vs \tau_k.
\end{equation}
From the definition of the multimode coherent state 
\begin{equation}
\ket{\Psi\ped{wp}} =  \exp(-\frac{1}{2}\int_{\mathbb R} \avs{\alpha(t)} \dd t)\sum_{n=0}^\infty\frac{1}{n!}\left(\int_{\mathbb R} \alpha(t)\adag(t)\dd t\right)^n\ket 0
\end{equation}
we can calculate~$\ev{\hat P_k(\vs \tau_k)}{\Psi\ped{wp}} = \eh{-\mu\ped{wp}}\prod_{j=1}^k \abs{\alpha(\tau_j)}^2$. Inserting this expression into the expression for $p\ped{wp}(t)$ yields the result
\begin{align}
p\ped{wp}(t)&=\eh{-\mu\ped{wp}}\sum_{k = 0}^\infty \int_{\mathbb R^k}\frac{1}{k!} p_{\text{click}}(t, \vs \tau_k) \prod_{j=1}^{k} \abs{\alpha(\tau_j)}^2\dd \vs \tau_k \notag\\
=\eh{-\mu\ped{wp}}&\sum_{k = 0}^\infty \int_{\mathbb R^k}\frac{1}{k!}\Bigg(\sum_{j=1}^{k} \eta \abs{\alpha(\tau_j)}^2 J(t-\tau_j)\prod_{l\neq j} \abs{\alpha(\tau_l)}^2\left(1- \eta \int_{\tau_l}^{t}J(t'-\tau_l)\dd t' \right)\Bigg)\dd \vs \tau_k\notag \\
&= -\eh{-\mu\ped{wp}}\pdv{t}\sum_{k = 0}^\infty \frac{1}{k!}\left(\int_{\mathbb R} \abs{\alpha(\tau)}^2\left(1- \eta \int_{\tau}^{t}J(t'-\tau)\dd t' \right) \dd \tau \right)^k\notag\\
&= -\pdv{t} \exp(-\eta  \int_{-\infty}^{t}(J*\abs{\alpha}^2)(t')\dd t').
\end{align}

\end{document}